# Magnetoresistance in Single Layer Graphene: Weak Localization and Universal Conductance Fluctuation Studies

Yung-Fu Chen, Myung-Ho Bae, Cesar Chialvo, Travis Dirks, Alexey Bezryadin, Nadya Mason

Department of Physics and Materials Research Laboratory, University of Illinois at Urbana-Champaign, Urbana, IL 61801-2902, USA

We report measurements of magnetoresistance in single-layer graphene as a function of gate voltage (carrier density) at 250 mK. By examining signatures of weak localization (WL) and universal conductance fluctuations (UCF), we find a consistent picture of phase coherence loss due to electron-electron interactions. The gate-dependence of the elastic scattering terms suggests that the effect of trigonal warping, i.e., the non-linearity of the dispersion curves, may be strong at high carrier densities, while intra-valley scattering may dominate close to the Dirac point. In addition, a decrease in UCF amplitude with decreasing carrier density can be explained by a corresponding loss of phase coherence.

## 1. Introduction

Studies of quantum interference are a key to understanding decoherence processes and scattering mechanisms in mesoscopic systems [1,2]. Graphene provides a particularly interesting subject for these studies because of its unique band structure, which is characterized by two points of valence band degeneracy (i.e., a "double-valley" degeneracy) (for recent reviews, see Ref. [3,4]). States with momenta near these degenerate points behave as chiral fermions; in this case, the role of spin is taken by a "pseudospin" which arises from the two equivalent sublattices of graphene's honeycomb structure [5]. The pseudospin is parallel to the momentum in one valley and anti-parallel to the momentum in the other; thus, the two valleys host quasiparticles having opposite chirality. It has recently been predicted [6] and shown experimentally [7-10] that the effects of chirality allow disordered single-layer graphene to demonstrate both weak localization (WL) and weak anti-localization (WAL) [11]. Weak localization occurs in a phase-coherent conductor when there is constructive interference between two time-reversed electron paths. This enhances elastic backscattering and is evident as a peak in magnetoresistance centered around zero magnetic field. However, in graphene, intravalley backscattering is suppressed because it requires the pseudospin to flip [12]. This suppression of backscattering reduces WL, and in fact shows up as WAL, which is evident as a zero-field dip in magnetoresistance [6]. Even though backscattering in graphene is suppressed, it can still occur if there is scattering off sharp defects: in this case intra- and inter-valley scattering can occur and restore WL. In addition, at higher energies the perfect alignment of the momentum and pseudospin is broken by "trigonal warping", or the asymmetric curvature of the dispersion away from the Dirac point, so the WAL effects should be reduced [6].

Previous work on WL in graphene has shown that the phase coherence length, $L_\varphi$, decreases with increasing temperature or decreasing carrier density, likely due to the effect of electron-electron interactions [7-10,13]. Fits to WAL have also been used to extract the elastic inter- and intra-valley scattering lengths. However, the gate-voltage trends in elastic scattering were previously difficult to determine, due to large scatter in the data and strong averaging over the signals [9]. Universal conductance fluctuations—another quantum interference effect that can be used to extract $L_\varphi$—have been seen previously in single-layer graphene, though their gate voltage dependence has not been compared quantitatively to WL results.

In this paper, we discuss magnetoresistance measurements on single-layer graphene in the context of WL and UCF. Fits to WL theory allow us to extract the carrier density-dependence of $L_\varphi$, as well as the carrier density-dependence of the inter-valley and intra-valley scattering terms. Our results seem consistent with the predicted strengthening of trigonal warping away from Dirac point. The values of $L_\varphi$ extracted from WL are compared to those obtained from autocorrelations of UCF (which is not affected by strength of elastic scattering) to obtain a consistent picture of the phase coherence in the system. We observe, and propose a mechanism for, a decrease in UCF amplitude with decreasing carrier density, an effect had been seen previously in bi- and tri-layer graphene but was not well understood [14].

**2. Sample Preparation and Measurements**

Our graphene samples were mechanically exfoliated onto highly doped Si substrates topped with 300 nm $SiO_2$. The thickness was determined by Raman and atomic force microscopy measurements. The electrodes were patterned by conventional electron beam lithography and electron beam evaporation of 3 nm Cr and 55 nm Au. The device shown in the inset of Fig. 1(a) consists of six electrodes on a piece of graphene, which allows measurements of longitudinal resistance $R_{xx}$ to determine 2D resistivity $\rho$, and Hall resistance $R_{xy}$ to determine carrier density $n_s$. All data presented in this paper are taken with the device shown in the inset of Fig. 1(a) (multiple samples showed similar effects, but only one was measured in detail). The distance $L$ between two longitudinal electrodes and the minimum width $W$ of the sample are 3.3 and 4.2 µm, respectively. Although the sample shape is not an ideal Hall bar geometry, the data still seem valid, as the overall longitudinal magnetoresistance has no slope (see Fig. 2), indicating no mixing from $\rho_{xy}$; similarly, the slope of the Hall resistance does not seem to be affected by $\rho_{xx}$ (which only adds an offset and contributions from universal conductance fluctuations). Measurements were performed in a Helium-3 cryostat using standard ac lock-in techniques: an ac current was applied through two end electrodes and the resulting longitudinal voltage or Hall voltage was recorded as a function of gate voltage $V_g$ (applied to the back of the doped substrate) or magnetic field $B$.

### 3. Results and Discussion

We determine the Dirac point using $R_{xx}$ and $R_{xy}$ as a function of $V_g$. A linear fitting of $R_{xy}$ vs $B$ gives the carrier density, via $dR_{xy}/dB = 1/n_s e$ (see Fig. 1(b)), where $e$ is the elementary charge. By extrapolating the carrier density vs $V_g$ plot to zero carrier density (see inset of Fig. 1(b)), we estimate the gate voltage at the Dirac point to be $V_{Dirac}$ ~ 64 V [15]. This value compares well to the turn-over point in a plot of $R_{xx}$ as a function $V_g$, as can be seen in Fig. 1(a). Also, as shown in Fig. 1(a), the gate voltage dependences of $R_{xx}$ and the residue $R_{xy}$ at $B = 0$ match very well, indicating the only effect related to the non-ideal Hall bar geometry in the sample is a field independent offset. The longitudinal resistance is also used to determine the mobility $\mu$ through the equation $\mu =$ $1/n_s\rho = L/n_s R_{xx} W$. At $V_g = 0$ (i.e., far from the Dirac point), $n_s \sim 5.9 \times 10^{12}$ cm$^{-2}$, $\mu \sim 1660$ cm$^2$/Vs, and the diffusion constant $D \sim 235$ cm$^2$/s.

Figure 2 shows magnetoresistance $R_{xx}(B)$ at different gate voltages. For all gate voltages, the resistance drops quickly when a small magnetic field is applied, resulting a sharp peak near zero magnetic field. This behavior is consistent with WL, showing that the carriers maintain their phase over many elastic mean free paths in the graphene. As the hole density decreases (i.e., as $V_g$ increases and approaches $V_{\text{Dirac}}$), the width of the WL peak increases, implying that the phase coherence length $L_\varphi$ of the holes decreases. In Fig. 2 it is also evident that at larger fields away from the WL peak the resistance increases with increasing magnetic field. This positive magnetoresistance becomes more pronounced near the Dirac point.

The graphene magnetoresistance data can be analyzed in terms of WL using theory described by Ref. [6] to find the resistance correction $\Delta\rho(B) = \rho(B) - \rho(0)$:

$$\Delta\rho(B) = -\frac{e^2\rho^2}{\pi h}[F(\frac{B}{B_\varphi}) - F(\frac{B}{B_\varphi + 2B_{iv}}) - 2F(\frac{B}{B_\varphi + B_{iv} + B_*})], \qquad (1)$$

where $F(z) = \ln z + \psi(\frac{1}{2} + \frac{1}{z})$, $\psi$ is the digamma function, and $B_{\varphi, iv, *} = \frac{\hbar}{4eD} L^{-2}_{\varphi, iv, *}$. Here, $L_{iv}$ is the elastic inter-valley scattering length. The third term in Eq. (1) predicts WAL, with $L_*$ related to the elastic intra-valley scattering length, $L_z$, and trigonal warping length, $L_w$, through the relation $L_*^{-2} = L_w^{-2} + L_z^{-2}$. From the scattering lengths we can extract the scattering rates $\tau^{-1}_{\varphi, iv, *}$ using the relation $L^2_{\varphi, iv, *} = D\tau_{\varphi, iv, *}$. Fits to Eq. (1) are plotted in Fig. 2 as thin solid curves. The theory fits the data well with $L_\varphi$, $L_{iv}$, and $L_*$ as three free parameters. In Fig. 3(a) and (b) we plot these characteristic length scales as a function of $V_g$. $L_\varphi$ is on the order of 1 μm, while $L_{iv}$ and $L_*$ are on the order of 0.1 μm in hole doped states; because $L_\varphi \gg L_{iv}$, the inter-valley scattering can act to restore the WL peak [9]. The decrease of $L_\varphi$ with decreasing carrier density has been observed before [7,9], and was explained by electron-electron interactions, specifically the inelastic Nyquist scattering of electrons off the fluctuating electromagnetic fields generated by all other electrons [9,16]. A similar decrease in $L_\varphi$ has also been seen as temperature was raised [17]. In addition, the decrease of $L_\varphi$ near the Dirac point may also be influenced by the formation of puddles of different types of carriers [9,18].

Figure 3(b) shows the intra-valley and inter-valley scattering lengths, demonstrating the weak increase of $L_*$ and $L_{iv}$ with decreasing hole density. This increase is expected for $L_*$ due to the influence of trigonal warping length, $L_w$, which should increase as the Fermi energy becomes close to the Dirac point as [6]

$$L_w^{-2} = \frac{\tau_w^{-1}}{D} = \frac{2\tau_0}{D}\left(\frac{\mu\varepsilon_F^2}{\hbar v_F^2}\right)^2 \qquad (2)$$

where $\tau_0$ is the momentum relation time, $\mu = \frac{\gamma_0 a^2}{8\hbar^2}$ is the structure constant, $\gamma_0 \sim 3$ eV is the nearest neighbor coupling, and $a = 0.142 \times \sqrt{3}$ nm is the lattice constant. The dotted line in Fig. 3(b) shows a plot of Eq. (2). As can be seen in the figure, at high carrier densities the calculated values of $L_w$ correspond closely to the extracted values of $L_*$. The correspondence between fitted $L_*$ and calculated $L_w$ at higher carrier densities ($V_g < 30$ V) implies that trigonal warping is the dominant chirality-breaking mechanism in this regime. At lower carrier densities the calculated $L_w$ diverges; in this case, the intra-valley scattering $L_z$ (extracted as dashed-dotted line in Fig. 3(b)) likely gives a cut-off length to $L_*$ and dominates the behavior. However, as the error estimation of $L_*$ shown in Fig. 3(b) is larger than the $V_g$ dependence of $L_*$, the effects of the trigonal warping and the intra-valley scattering described above may need more accurate measurements to verify. In Fig. 3(b) it can also be seen that the inter-valley scattering length $L_{iv}$ has weak gate-voltage dependence, and seems to track $L_*$. The cause of this behavior is unknown, and may imply that effects in addition to trigonal warping are influencing the elastic scattering. In particular, it is possible that some of the positive magnetoresistance (which is used to extract $L_*$ and $L_{iv}$) is due to charge inhomogeneity in the system [19]. Although this effect should be less relevant than WL/WAL at low temperatures, results on bilayer samples show that charge inhomogeneity effects may increase with decreasing carrier density [20].

Because the WL fits to $L_\varphi$ involve consideration of elastic scattering terms, it is valuable to compare these results to a method of obtaining $L_\varphi$ that does not involve other scattering terms. To do this, we extract $L_\varphi$ from UCF. UCF occurs in disordered, mesoscopic conductors where the carrier transmission is due to interference of multiple, complicated paths which accumulate a random phase when a parameter such as magnetic field is tuned. The resultant conductance fluctuations are reproducible and aperiodic, and have amplitudes of order $e^2/h$. UCF is evident in Fig. 2 as $B$-field-dependent fluctuations of $R_{xx}$. We separate the UCF from the WL signals by defining

$$G_{fluc} = R_{xx,data}^{-1} - R_{xx,WLfit}^{-1}, \qquad (3)$$

Figure 4(a) shows the UCF as a function of $B$ for different $V_g$, where the magnitude is on the order of $e^2/h$. The reproducibility of the fluctuations is shown in the inset of Fig. 4(a); UCF from both the forward and backward magnetoresistance sweeps are nearly indistinguishable. The frequency and amplitude have a clear dependence on $V_g$ (hole density). We calculate the autocorrelation $G_{fluc}(B)G_{fluc}(B+\Delta B)$ of each UCF curve to find the characteristic frequency $B_c$ and variance $G_{fluc,var} = \langle(G_{fluc} - \langle G_{fluc}\rangle)^2\rangle$, where $\Delta B$ is the magnetic field lag, $B_c$ is the half width at half maximum of the autocorrelation and $G_{\text{fluc,var}}$ is the autocorrelation at $\Delta B = 0$. The extracted phase coherence length, $L_\varphi = (2.4h/eB_c)^{1/2}$, is plotted in Fig. 4(b), while the root mean square fluctuation amplitude, $G_{\text{fluc,rms}} = (G_{\text{fluc,var}})^{1/2}$, is plotted in Fig. 4(c). As can be seen in Fig. 4(b), $L_\varphi$ from the UCF analysis and from the WL fitting agree well with each other, validating the extracted values of $L_\varphi$. The similar trend of $L_\varphi$ decreasing with carrier density shows that the origins of the electron phase breaking process is the same for both phenomena.

An unusual aspect of the UCF is the suppression of fluctuations as $V_g$ moves toward the Dirac point, as can be seen in Figs. 4(a) and 4(c). Although this phenomenon has been recently shown in bi-layer and tri-layer graphene [14], it has not been reported before in single-layer devices. While the origin of this suppression was not previously clear, it likely occurs because $L_\varphi$ decreases in a device where $L_\varphi$ is smaller than the channel length. In this case, we follow the argument in Ref. [2], which considers a 1D strip of the device ($W < L_\varphi$, $L > L_\varphi$) as composed of $N = L/L_\varphi$ series resistors. The 1D strip fluctuation amplitude is $\delta g_{1D} = N^{-3/2}\delta g_0 = (L_\varphi/L)^{3/2}\delta g_0$, where $\delta g_0 \sim e^2/h$ is the fluctuation for each resistor. A 2D conductor such as our sample ($W > L_\varphi$, $L > L_\varphi$) is composed of $N = LW/(L_\varphi)^2$ squares, where each square has size $(L_\varphi)^2$ and fluctuation $\delta g_0$. The entire sample then has fluctuation amplitude $\delta g_{2D} = (W/L)N^{-1/2}\delta g_0 = L_\varphi(W^{1/2}/L^{3/2})\delta g_0$. The fluctuation amplitude should thus scale with $L_\varphi$. As can be seen by a comparison of Fig. 4(b) and 4(c), when $L_\varphi$ decreases by a factor of two over 55 V, the fluctuation amplitude decreases by about the same factor over the same gate voltage range, consistent with expectations.

## 4. Summary

In summary, we have examined weak localization and universal conductance fluctuations in single-layer graphene as a function of carrier density. The magnetoresistance curves fit well to the predicted theory of WL/WAL in graphene. UCF data was analyzed via autocorrelations to determine the gate voltage dependence of the fluctuation frequency and amplitude. The phase coherence lengths extracted from WL and UCF correspond well, validating the fitting procedures for both measurements and showing that the origins of the electron phase breaking process is the same for both phenomena. We also explained a decrease in UCF amplitude with decreasing carrier density as due to a corresponding loss of phase coherence. From the WL data, we determined the gate-dependence of the elastic scattering terms and suggested that the trigonal warping effect is strong at high carrier densities. However, the similar gate trends of inter- and intra-valley scattering may imply that positive magnetoresistance due to charge inhomogeneity is also relevant.

The authors wish to thank Peter Abbamonte, Norman Birge, and Michael Stone for useful discussions. This work was supported by the U.S. Department of Energy, Division of Materials Sciences under Award No. DE-FG02-07ER46453, through the Materials Research Laboratory and Center for Microanalysis of Materials (DE-FG02-07ER46453) at the University of Illinois at Urbana-Champaign.

References:

[1]   C. W. J. Beenakker and H. Vanhouten, Solid State Physics-Advances in Research and Applications **44**, 1 (1991).

[2]   S. Datta, *Electronic transport in mesoscopic systems* (Cambridge University Press, 1997).

[3]   A. K. Geim and K. S. Novoselov, Nature Materials **6**, 183 (2007).

[4]   A. H. Castro Neto, F. Guinea, N. M. R. Peres, K. S. Novoselov, and A. K. Geim, Reviews of Modern Physics **81**, 109 (2009).

[5]   Y. S. Zheng and T. Ando, Physical Review B **65**, 245420 (2002).

[6]   E. McCann, K. Kechedzhi, V. I. Fal'ko, H. Suzuura, T. Ando, and B. L. Altshuler, Physical Review Letters **97**, 146805 (2006).

[7]   S. V. Morozov, K. S. Novoselov, M. I. Katsnelson, F. Schedin, L. A. Ponomarenko, D. Jiang, and A. K. Geim, Physical Review Letters **97**, 016801 (2006).

[8]   X. S. Wu, X. B. Li, Z. M. Song, C. Berger, and W. A. de Heer, Physical Review Letters **98**, 136801 (2007).

[9]   D. K. Ki, D. Jeong, J. H. Choi, H. J. Lee, and K. S. Park, Physical Review B **78**, 125409 (2008).

[10]  F. V. Tikhonenko, D. W. Horsell, R. V. Gorbachev, and A. K. Savchenko, Physical Review Letters **100**, 056802 (2008).

[11]  F. V. Tikhonenko, A. A. Kozikov, A. K. Savchenko, and R. V. Gorbachev, Physical Review Letters **103**, 226801 (2009).

[12]  P. L. McEuen, M. Bockrath, D. H. Cobden, Y. G. Yoon, and S. G. Louie, Physical Review Letters **83**, 5098 (1999).

[13]  D. Graf, F. Molitor, T. Ihn, and K. Ensslin, Physical Review B **75**, 245429 (2007).

[14]  N. E. Staley, C. P. Puls, and Y. Liu, Physical Review B **77**, 155429 (2008).

[15]  The device was unfortunately killed as we tried to sweep gate voltage above Vg ~ 68 V.

[16]  N. O. Birge and F. Pierre, in *the Summer School on Quantum Phenomena of Mesoscopic Systems*, edited by B. L. Altshuler, A. Tagliacozzo and V. Tognetti, Varenna, Italy, 2002).

[17]  In general, $L_\varphi$ is expected to increase as temperature is lowered, while $L_*$ and $L_{iv}$ should be largely temperature independent. Although our preliminary data support this, the sample was unfortunately destroyed before systematic temperature data could be obtained.

[18]  J. Martin, N. Akerman, G. Ulbricht, T. Lohmann, J. H. Smet, K. Von Klitzing, and A. Yacoby, Nature Physics **4**, 144 (2008).

[19]  S. Cho and M. S. Fuhrer, Physical Review B **77**, 081402(R) (2008).

[20]  Y.-F. Chen, M.-H. Bae, C. Chialvo, T. Dirks, A. Bezryadin, and N. Mason, condmat, 0906.5090.

[21]  K. S. Novoselov, A. K. Geim, S. V. Morozov, D. Jiang, M. I. Katsnelson, I. V. Grigorieva, S. V. Dubonos, and A. A. Firsov, Nature **438**, 197 (2005).

[22]  Y. B. Zhang, Y. W. Tan, H. L. Stormer, and P. Kim, Nature **438**, 201 (2005).

Figure Captions:

Figure 1. Graphene device characterization. (a) Longitudinal resistance $R_{xx}$ (solid curve) and the residue $R_{xy}$ at $B = 0$ (solid squares, data from (b)) as a function of $V_g$ at 250 mK. Inset: Optical image of the measured device. (b) Hall resistance for different backgate voltages, $V_g$ ($V_g$ varies from -40 V to 40 V in steps of 10 V). All data is taken at 250 mK. The slope of the Hall resistance is used to determine the carrier density at each $V_g$. Inset: Carrier density (in units of $10^{12}$ cm$^{-2}$) as a function of $V_g$. The positive carrier density indicates that the carriers are holes. The linear fit in the inset shows that the backgate-induced change in carrier density is $9.26 \times 10^{10}$ cm$^{-2}$V$^{-1}$, in good agreement with $7.19 \times 10^{10}$ cm$^{-2}$V$^{-1}$ estimated from the device's geometric capacitance [21,22].

Figure 2. Longitudinal magnetoresistance and WL theory. $R_{xx}(B)$ at 250 mK for different $V_g$ ($V_g$ varies from 50 V to -5 V in steps of 5 V). A WL peak near zero magnetic field is observed at all gate voltages. Thin solid lines show the fit to the graphene WL theory proposed by McCann *et al.* [6]. The inset shows the blow-up view of the curves for $V_g = -5$ V and 0 V (the axis labels are the same as those on the main plot).

Figure 3. Characteristic lengths obtained from WL fitting of longitudinal magnetoresistance in Fig. 2. (a) Phase coherence length, $L_\varphi$, as a function of $V_g$. The dotted line is the linear fit. (b) Elastic scattering lengths as a function of $V_g$. The dashed line is a linear fit of $L_*$, to demonstrate the trend of increasing length with $V_g$. The dotted line shows trigonal warping $L_w$ calculated from Eq. (3). The dashed-dotted line is $L_z$

obtained from $L_z^{-2} = L_\phi^{-2} - L_w^{-2}$ (using $L_*$ extracted from data and the calculated $L_w$; for simplicity, $L_z$ is assumed to be independent of gate voltage).

Figure 4. UCF fluctuations $G_{fluc}$ as a function of magnetic field $B$ at different $V_g$ ($V_g$ varies from -5 V to 50 V in steps of 5 V). The data is obtained by subtracting the solid-line fits from the data in Fig. 2. An offset of $2e^2/h$ is applied between adjacent curves for clarity. Inset: $G_{fluc}$ from both forward and backward magnetoresistance sweeps for $V_g$ = -5 V in Fig. 2(a) (red solid and blue dotted lines, respectively), demonstrating that the UCF is reproducible. (b) Comparison of $L_\varphi$ obtained from UCF (blue empty squares) and WL (red solid squares) analyses of longitudinal magnetoresistance (c) UCF amplitude (root-mean-square of $G_{fluc,rms}$) vs $V_g$.

Figure 1

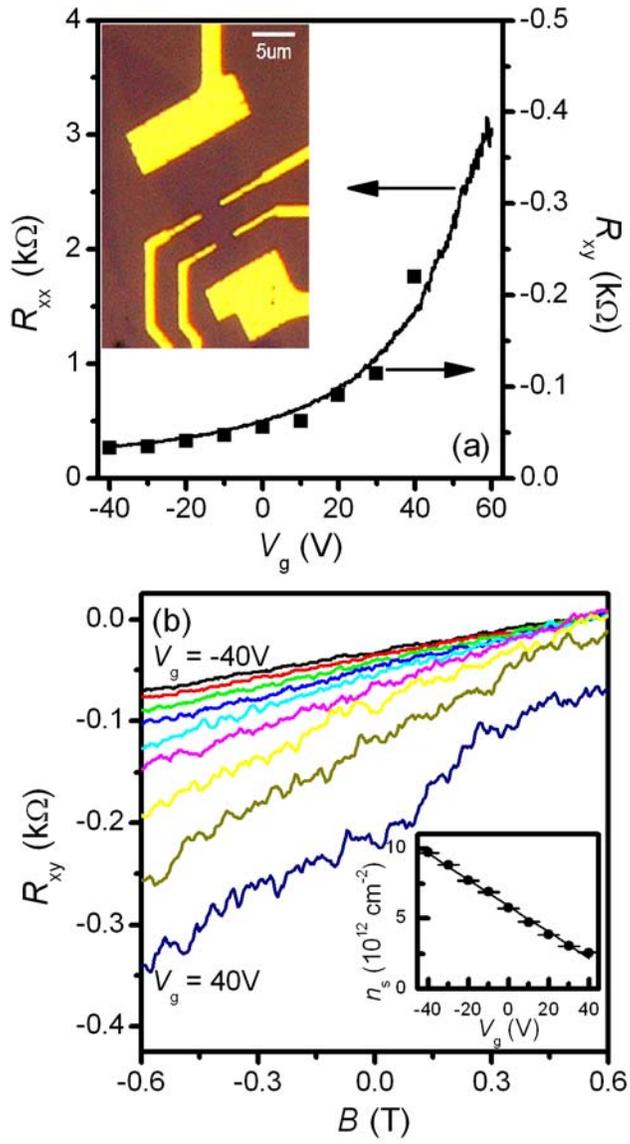

Figure 2

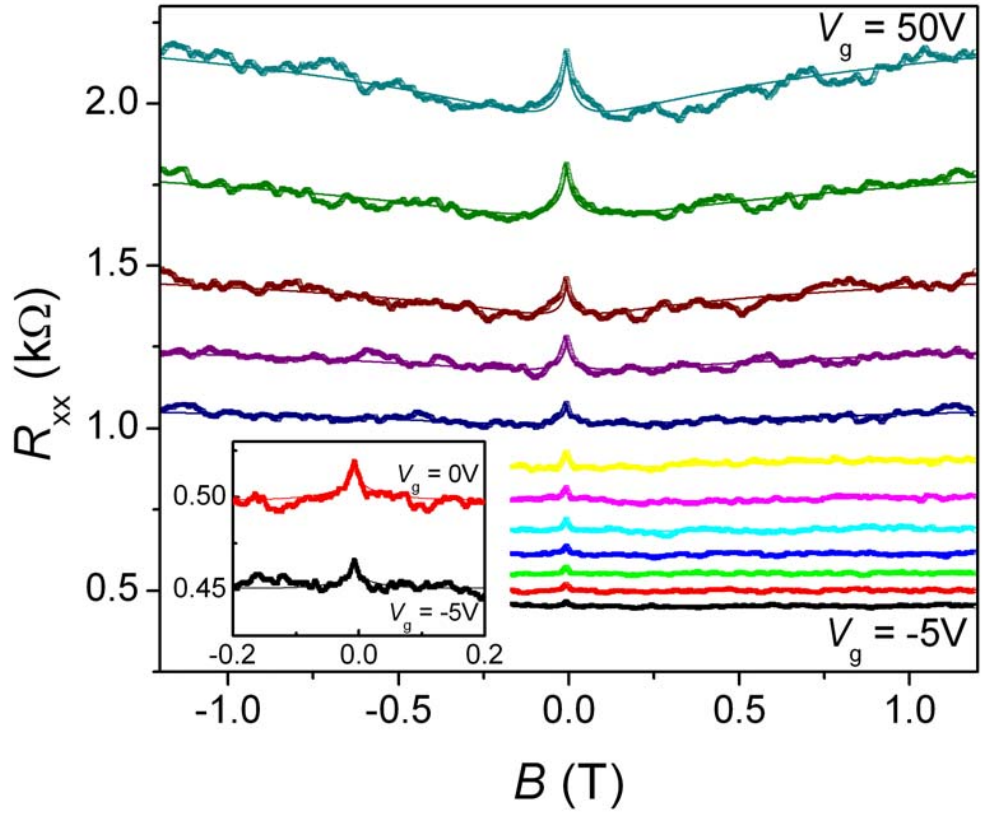

Figure 3

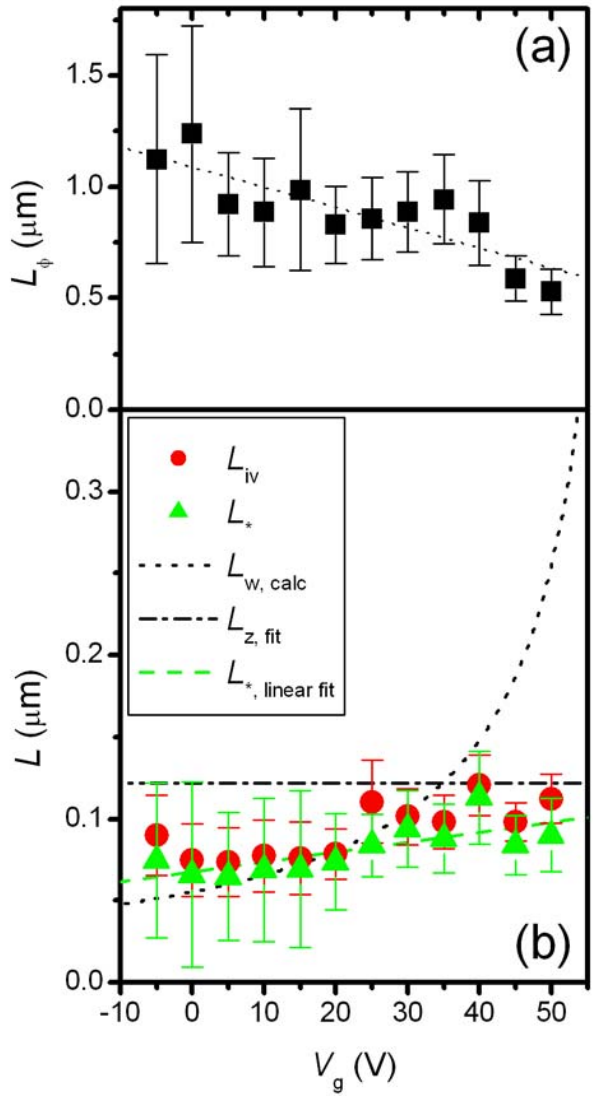

Figure 4

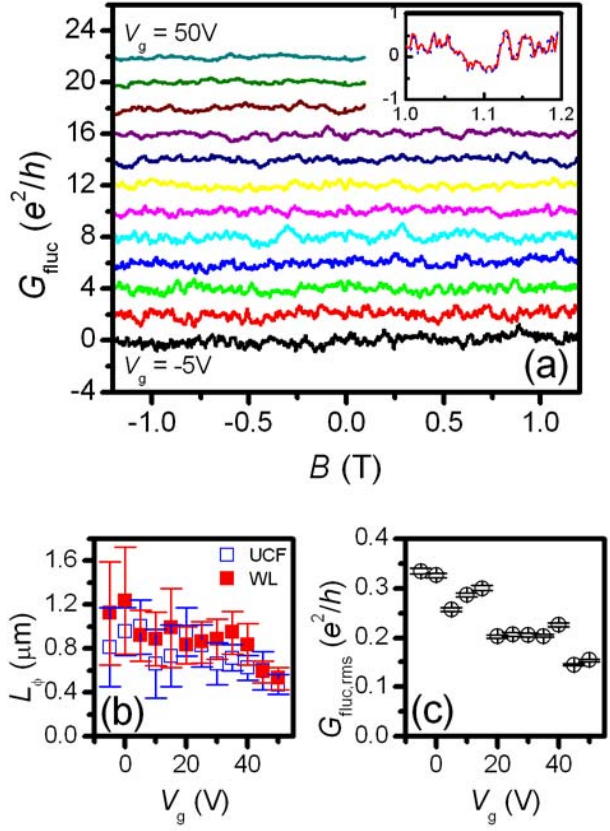